\input{aipcheck}

\documentclass[
    ,final            
  ]
  {aipproc}

\layoutstyle{6x9}

\begin{document}

\title{First Stars IV: Summary Talk}

\classification{97.20.Wt, 98.54.Kt, 98.62.Ra, 98.80.-k}
\keywords     {First stars and galaxies; cosmic reionization; black holes}

\author{Andrea Ferrara}{
  address={Scuola Normale Superiore, Piazza dei Cavalieri 7, 56127  Pisa, Italy}
}




\maketitle


The fourth conference of the ``First Stars'' series was held in Kyoto. In addition to the beauty and splendor
of the city, this choice also clearly contains a recognition of the outstanding contributions brought by Japanese
scientists in the area of (first) star formation. This tradition dates back to the pioneering work and gigantic 
achievements of Prof. Hayashi, which have been reviewed along with many important facts of his career by
a number of very nice talks during the first day of the conference and are reported elsewhere in these proceedings.

To put this conference in perspective against the previous ones (First Stars I-III) held every four years starting
from 1999 (First Stars IV was posponed by one year following the tremendous shock of the earthquake/tsunami
that has devastated  the east coast of Honshu in 2011) a few general comments might be in order. After 13 years
the physics of first star formation has been greatly clarified and much more sophisticated models and predictions
have been developed. Numerical simulations have now included a number of ingredients such as enlarged chemical
networks, radiation transfer and pressure, dust grains, magnetic fields, accretion disks, ionization fronts which 
on the one hand make the various treatments more realistic; on the other hand, they have increased the complexity of
the problem to a level at which predictions depend in an unpredictable manner on the adopted  physical coefficients, 
initial conditions and numerical resolution. 

A common way to proceed in many physical areas in order to fix parameters that cannot be obtained from 
ab initio calculations is to get guidance from experiments. Unfortunately, the lack of positive detection and hence 
of direct data on the nature of the first stars appears at the moment the major difficulty in the field. In spite of the
fact that the search for metal-free stars has continued for decades in the halo of the Milky Way, none has been found;
the elemental abundances of metal-poor stars in the Local group have yielded only very mild constraints on the
IMF of the early generation of stars, incidentally often in conflict with theoretical expectations;  the spectra of intermediate 
and high redshift galaxies do not indicate the expected signatures of metal free stars in their SEDs; direct detection of PopIII 
clusters will probably remain unfeasible even with the forthcoming James Webb Space Telescope. We should not be discouraged
by these apparent failures; rather, we should be aware that we are trying to push the frontiers of knowledge beyond a very  distant
cosmic frontier and it is of foremost importance to keep studying, collecting information and new data which will eventually be
used to recompose the puzzle when, hopefully soon, times will be ripe. 

The main topics discussed at the conference can be somewhat arbitrarily grouped into four areas: Formation and evolution 
of the first stars and their local relics; First stars detection through their most luminous and spectacular evolutionary phases 
(supernovae and Gamma-Ray Bursts) and in the most distant galaxies; Cosmic reionization; First black holes. The main 
results presented are discussed in the following Sections according to this agenda.

\section{First Stars: Formation and Evolution}
An important point that has been highlighted at this conference is that a solid understanding of the final mass and properties
of the first stars has to pass through a correct description of the evolution of the accretion disk resulting from the initial collapse
of the rotating proto-cloud. Such disk, feeding the central seed protostar, is rotationally supported but can become Toomre unstable
due to the cooling provided by H$_2$ lines, collision-induced emission, and H$_2$ dissociation, eventually leading to the formation of one 
or more fragments in the central region of the disk.  The disk continues to accrete matter from the surroundings at a faster rate than 
the protostar feeding; at the same time the fragments migrate towards the center due to gravitational torques. It is estimated that 
half of them merge with the central protostar, likely causing the interesting phenomenon of accretion bursts; the remaining ones are
sling-shot out of the formation site and perhaps in the intergalactic medium. The latter possibility is interesting as it enables to make 
relatively constrained predictions on the number of brown dwarfs/very low mass stars at $z=0$, that can be tested against 
observations. Uncertainties remain on whether these fragments are bound and hence able to survive as they are expelled. 

The above scenario is however far from being unanimously accepted. Some groups, for example, have stressed the 
key role of turbulence, which might prevent disk formation by providing pressure support arising from an increased core temperature
induced by vorticity dissipation.
This picture would then dramatically differ from the accreting disk one introduced above, with presumably widely different 
expectation values for the final mass of the star. The point is that turbulence effects might have been so far underestimated by 
numerical experiments that do not resolve the Jeans length at sufficiently high  resolution, thus possibly explaining the 
different outcome. It will be necessary to investigate in more detail the sources, spectrum and dissipation rate of turbulence 
under these conditions before we can draw a conclusion on its role.  

A final point is that at this stage 3D simulations are unable to give a firm prediction on the mass of the first stars. 
A major technical problem is that, having abandoned the necessary, but physically shaky, sink particle technique in favor of a full
hydrodynamical description, simulations can follow the above process only for a timescale limited to $< 10$ yr. The longer term
evolution can instead be tracked by 2D simulations including radiative transfer (and qualitatively confirmed by ad hoc 3D 
lower resolution simulations including similar physics) which found that the final mass of the star is $< 50 M_\odot$. Such value arises from 
the quenching of accretion from the disk caused by the expansion of the HII region powered by the ionizing photons produced by the star as soon as its
mass becomes larger than $\approx 10 M_\odot$.  The above upper limit has strong empirical implications: it would imply, among other things, 
that no Pair Instability SuperNovae (PISN) should occur at high redshifts. This result also poses two urgent issues to the 3D 
simulators: (i) it shows that long integration times ($>10^5$ yr) are required to follow the entire accretion history; (ii) it questions
disk fragmentation in the presence of the ionizing radiation from the protostar which might stabilize the disk by heating it.  

Additional interesting effects can arise if the dark matter is annihilating (again heating/stabilizing the disk) and/or a magnetic
field is present. While at first core formation the magnetic energy is subdominant with respect to the kinetic one, 
the rapid growth induced by vorticity generated by shocks or chemo-thermal gradients might lead to field amplification on
a timescale of $10^{-4}$ free-fall times, \emph{if} a small-scale turbulent dynamo operates -- which has yet to be demonstrated.

A crucial and debated issue concerns the structural differences between the first stars and the present-day ones, a research 
area often referred to as the ``PopIII-PopII Transition''. This problem is usually studied through semi-analytical or numerical 
implementations of the phase diagram $T-n$ of a collapsing proto-cloud that embeds the various cooling processes
concurring to gravitational energy dissipation. Once this has been worked out, the thermodynamic properties and 
equation of state determine the cloud fragmentation condition (and by extrapolation the final mass of the star), i.e. fragmentation
occurs when the Jeans mass stops decreasing with increasing density, or the adiabatic index $\gamma > 4/3$ for a  spherical geometry.
While there is good agreement on the usefulness of this method, a debate has arisen in the last years on which cooling channel, i.e. 
dust grains or gas phase C and O excitation, was the key one to fix the critical metallicity. As gas phase-dominated fragmentation 
cannot work below a metallicity $Z = 10^{-3.5} Z_\odot$, the very recent discovery of the SDSS J102915+172927 star with total 
metallicity $Z = 10^{-4.35} Z_\odot$ has possibly set an end to the discussion in favor of the dust cooling mode and a critical 
metallicity of $Z= 10^{-5 \pm 1} Z_\odot$. 
   
Understanding the critical transition might enable us to make detailed predictions on a number of important issues as, e.g.
the fraction of galaxies purely made of PopIII stars and their redshift frequency, the fraction of core collapse supernovae
vs. hypernovae, the rate of PopIII GRBs at the earliest epochs. The relative star formation rate in PopIII and PopII stars might
also help us interpreting rare and puzzling phenomena as the recently discovered class of very luminous supernovae, 
represented by the SNIIn found at $z=2.05$ and reported at this conference. 

In spite of this huge and noticeable efforts we are still far from having a sound understanding of the first stars. Fortunately,
stellar archeology appears to be able to shed some light on this difficult matter. Here are a few facts that we have learned 
through this technique. First, no truly metal-free stars and no star with total metallicity $Z < 10^{-4.35} Z_\odot$ have  
been detected in >30 years. Second, only a handful of Ultra Metal Poor stars with [Fe/H] $< -5$ have been found. 
The Metallicity Distribution Function (MDF) rapidly rolls-off at [Fe/H]$< -3.5$, whereas the abundance of C-rich ([C/Fe] $>1$) stars 
increases at low [Fe/H]. The Lithium plateau melts-down and the scatter of measurements increases at low [Fe/H], something
often interpreted as a signature of stellar rotation. Also, the nature of $r$-processes (i.e. [Sr/Ba]) changes at [Fe/H]$< -3.5$. Finally,
the fraction of Extremely Metal Poor binaries is as high as 10\%. It is at present unclear how to put all this data in a coherent
framework. Certainly these facts should embed precious information about the formation process, IMF, transition and evolution
of the first stellar generation, but extracting and filter out such relevant information will take several more years of investigations.  

Maybe we are just looking in the wrong place by mining the Galactic halo and we should look at the dwarf satellites of the Milky Way, 
including the mysterious Ultra Faint Dwarfs (UFD) discovered by SDSS in the last few years. We have already learned a lot from the initial
inspection of these objects. For example, the dwarf galaxies metallicity-luminosity relation and MDFs all require an epoch of
pre-enrichment building up a common metallicity floor. The [$\alpha$/Fe] ratio at fixed [Fe/H] in dSphs appears to be lower 
than in the MW halo stars indicating that outflows must have been very important for these low-mass objects; also it has been
shown that environmental effects (tidal, ram-pressure stripping) play only a minor role in the overall evolution of these systems.
UFD are particularly interesting as they could be leftovers of the earliest galaxies, probably mini-halos; they are the best candidates 
to search for PopIII stars (some tantalizing indications in this sense have been presented based on the analysis of the Hercules UFD data).
An even more intriguing link of these peculiar objects is with the recently discovered C-enhanced DLAs ($z=2-3$),  
that have been interpreted as ``failed'' UFDs, that is galaxies in which the star formation efficiency was too low to get rid of their gas.

\section{First Stars Detection}
As already pointed out, we are in strong need of positive detections of truly pristine stars to guide our
understanding and theoretical insights. However, this has proven so far extremely challenging both
locally, and at intermediate and highest redshifts. The reason might well be that, albeit by construction
the first stars must have been metal-free according to the very well tested Big Bang Nucleosynthesis theory,
their formation phase might have been extremely short, thus leaving behind almost no traces. Theoretical models however
inform us that although at rates a few orders of magnitude less than PopII stars,  PopIII star formation can have 
continued unhindered for several billion years in the pristine, low density spots perhaps corresponding to the
peripheries of galaxies. This motivates us to continue the searches with any possible mean. 

The most obvious way to detect PopIII stars is to trace their final evolutionary phase when they explode as
supernovae. We have already mentioned that given the recent developments we might not expect PISNs 
at high redshifts. It has to be kept in mind, though, that many physical aspects of the problem are only started to be appreciated and understood; hence the situation might change again and
it is a good policy to keep an open mind when we design our experiments/observations. Both theory and observations
seem to favor (consistent with the recent findings of the upper limit of $50 M_\odot$ reported above) hypernovae
as the dominant enrichment agent of the metal-poor halo stars, as deduced from their elemental abundance patterns.
We should be rather satisfied also with the fact that both pre-SN evolution and the explosive nucleosynthesis/fate are now
well understood along the entire WD/NS/BH/PISN/BH sequence, allowing to reliably predict metal ejection.
Uncertainties (like rotation, mass loss and mixing) are still present, but we are making very rapid progresses
in this area and we do expect even more robust results shortly.

In principle high-$z$ SNe constitute also the most promising way to spot the formation sites of PopIII stars. Because
of time dilation of the light curves, their identification will require a careful design of the monitoring-and-revisiting
observational strategy. These difficulties would be partly simplified by the suggested use of the shock-breakout detection.
The good news are that fully-fledged and reliable PopIII light curves have now been computed. Based on this 
data, results have been presented claiming that JWST might see PISNs up to $z=15-20$, a remarkable achievement
paving a new road to explore the first cosmic star formation phases. More indirectly, one could search in reasonably
luminous H-cooling halos (but how many of them are still metal-free?) which could be prototype PopIII galaxies.
Their detection with JWST would require that in these systems at least 10\% of the baryons are turned into stars, 
an efficiency probably too much on the high side to be the rule. Lensing by foreground matter could decrease
this requirement to 0.1\%. The open question remains of how to disentangle PopIII galaxies from normal ones with
photometry only. 
 
A second, slightly more convolved, possibility is to use Gamma-Ray Bursts (GRB). Long GRBs are connected with hyper/supernovae
as predicted by the collapsar model. The GRB record holders at present are the spectroscopically confirmed 
GRB090423 at $z=8.2$, and the more distant but photometric only GRB090429 at $z=9.4$. GRBs are intrinsically much  
brighter than galaxies at the same redshift; hence they act as luminous signposts of star formation activity otherwise 
undetectable. Indeed, their detection can provide an independent method to determine the cosmic star formation history.
By comparing GRB-determined rates with the ones derived from high-$z$ galaxy surveys as the HUDF, there is a tendency
for the former to be a factor of several higher than the latter, maybe as a consequence of the fact that a large fraction
of the star formation in the early universe occurs at levels below detection or (less probably) in regions that are heavily 
dust-obscured. The GRB host galaxies gas metallicities appear to be  -- at least at intermediate redshifts -- lower than the
mean of the galaxy population, but this aspect is still heavily debated. What can we learn on PopIII stars from GRBs is at
the moment not completely clear. One major uncertainty is whether PopIII GRBs can get rid or emerge from the massive 
envelope of their progenitor: present studies do not have a unique answer to this question, some of them appealing
to the presence of a binary companion to make the job easier. All these aspects need to be clarified before we can 
safely use GRBs to infer the physical properties of the first/progenitor stars, 

Primeval galaxies forming during the first billion year after the Big Bang in principle should represent optimal
locations to search for the first stars, in addition to constitute a subsample of the reionization sources, as 
we discuss later.  These objects are usually found through two different complementary techniques: (i) the
so-called ``drop-out'' technique in which an object disappears from photometric bands with central wavelength
shorter than the rest-frame Ly$\alpha$ line due to interstellar and intergalactic scattering of the line by neutral 
hydrogen; (ii) narrow-band filters tuned on the rest-frame Ly$\alpha$ line frequency. Objects found with the
first technique are dubbed Lyman Break Galaxies (LBGs) while the latter are called Lyman Alpha Emitters (LAE).
Recent data reported at the conference witnesses the tremendous advances made in the last few years that
have allowed us to spectroscopically confirm galaxies at redshift $ \approx 7$ and to build the galaxy Luminosity Function
(LF) up to $z\approx 9-10$ using LBG candidates. These techniques, and particularly (i), suffer of many uncertainties
which should be always kept in mind when handling this type of data. Among these are the fact that interlopers can 
resemble high-$z$ galaxies at low S/N, uncertainties related to the size/surface brightness distribution leading to 
sample incompleteness, the selection efficiency, the estimate of the cosmic variance and dust corrections. 
In spite of these warnings a relatively solid experimental body of data emerges out of which we can build LFs with some confidence
and infer the evolution of the parameters of the Schechter function template, indicating a steepening of the faint end
to slopes close to $-2$. If indeed the LF at high $z$ can still be modelled correctly through
such an empirical functional form remain to be ascertained. 

Until recently there was hope to be able to constrain the presence of PopIII stars in high $z$ galaxies via their spectral 
UV slope $\beta$ (the specific flux shape, $f_\lambda \propto \lambda^\beta$). Very blue colors ($\beta \sim -3.5$) were initially 
found, which could be reproduced only with the concomitant presence of young metal-free stars and large values of the escape 
fraction. This evidence has been recognized now as a spurious one and at this meeting much more standard values of $\beta \approx -2.0$, with a mild increasing trend 
with time, have been reported. Even the highest redshift galaxies seem to be dominated by relatively evolved stellar 
populations with no indication of first stars features, hence stiffening the problem of finding these elusive early stars. 
In addition to the LFs themselves, a lot of information can be obtained from the measurement of their time integral, i.e. 
the stellar mass function, at different redshifts.

\section{Cosmic reionization}
One of the most intriguing questions about the first stars is if and to what extent they have contributed to the
observed reionization of cosmic hydrogen and helium. Their structural properties (low opacity, high surface 
temperatures and mass) are such that they are expected to overshine by a factor $6-25$, depending on their IMF,
the ionizing photon production by PopII stars. Thus, in principle, they are optimal reionization candidate sources.

Our understanding of reionization has certainly become much more extensive in the last few years thanks
to a wealth of experiments that have allowed to build data-constrained reionization models. The overall
features of reionization are described in terms of overlapping ionized bubble structures, moving initially
from the high density peaks (a density-biased evolution) and later propagating into the voids to ionize the 
remaining neutral islands. Only about 3 ionizing photons per HI atom per Hubble time are available at $z=6$,
a property that is often referred to as a ``starving reionization'' having to rely on a number of photons barely
sufficient to balance recombinations in the IGM. A very important point that has been re-emphasized at the conference
is that reionization by $z=6$ requires the emissivity $d\epsilon/dz = d(J_\nu /\lambda_{mfp})/dz > 0$, i.e. to
increase at earlier times. This trend could be caused by an increase of the escape fraction, $f_{esc}$, by a larger
star formation efficiency or by a shift to a top-heavy IMF towards high redshift; at the moment it is unclear what factor 
is the dominant one.  However, there is good consensus that reionization must be driven by sources that are (much) fainter
that those currently part of LFs built through deep surveys as the HUDF: a concordance statement perhaps would
indicate that at $z=7$, more that  $50$\%  of the total ionizing photon budget comes from dark matter halos with 
mass $< 10^9 M_\odot$. These sources will present a serious detection challenge even for JWST.  Fortunately, though,
the number of possible experiments to investigate reionization is increasing rapidly and we have great hopes in the
success of HI 21 cm redshifted line experiments which should be able to obtain a detailed tomography of the 
HI distribution in the reionization epoch. Also, very interesting constraints might come from the Near Infrared Background
and the kinetic SZ, which have already provided some unique insight on the reionization history.
Finally, the search for low-$z$ analogs of the reionization sources would allow us to better understand the link between
the observed properties (as e.g. the Ly$\alpha$ line luminosity and profile) and the morphology, star formation rate and $f_{esc}$. Very promising
preliminary results have been presented which are likely to open a new important area of investigation.   

Although the reionization history seems to be significantly bound, a strong effort is still required to understand
the relative role of the sources (including the more exotic ones as dark matter annihilation/decay) and, even more importantly, 
the feedback effects that reionization has produced on galaxy formation history. 
These feedbacks are usually classified as mechanical (related to supernova explosions), radiative (produced by soft UV and
ionizing radiation) and chemical (due to the presence of metals and dust grains). Assessing their importance at various times
and for different galaxy mass ranges is the final goal of the current studies. This task is far cry from being simple though,
as a result of the large number of spatial and time scales involved, and of the complexity of the problem. A few key points 
that need urgent attention have been outlined: it is mandatory to better understand the gas/metal ejection efficiencies of small 
and large galaxies and the level and intermittency of turbulent energy injection and dissipation in the interstellar/intergalactic 
medium; is the radiative feedback in relic HII regions positive or negative ? What is the level of  suppression of galaxy formation 
below the Jeans filtering  mass and is there a sharp boundary between the two regimes or rather a gentle decrease of the baryonic 
fraction ? How can we quantify more precisely the effects of the global Lyman-Werner background sterilization of mini-halos ?
These are fundamental questions to which we have only partial and qualitative answers at present; nevertheless, it is hopeless to 
understand reionization in full detail lacking a comprehension of these physical aspects. 
 
\section{First Black Holes}
The formation of the first black holes is a fascinating topic central to many aspects of galaxy formation and
reionization. The presence of several Super Massive Black Holes (SMBH) with masses in excess of $10^9 M_\odot$
at $z>6$ and the recent discovery of a new one at $z=7.085$ ($t=0.77$ Gyr) imply that black hole formation/growth
has to: (a) start early (b) work at (super-)Eddington accretion rates at all times. These two facts already favor seed
BH masses $>400 M_\odot$; in addition a Eddington accretion of all the formed seeds might cause serious conflicts with the
upper bounds set by the unresolved fraction of the X-ray background.

BH seeds can be produced either as a final product of the evolution of stars more massive than
$30-40 M_\odot$ or by the direct collapse of a gas cloud. In the first case one has to make sure that the BH can grow rapidly
by providing a continuous gas supply, avoiding radiation feedback from the growing hole depressing the accretion rate, prevent
ejection from the parent halo; in addition one has to be careful not to overproduce $\approx 10^6 M_\odot$ holes at any
observable redshift. These problems might look as show-stoppers, but the alternative route has to cope with even more
critical issues. In order to produce a BH by direct gas phase collapse, fuel must be driven in rapidly (i.e. a deep potential is needed), 
angular momentum must be transferred efficiently; it is necessary to avoid fragmentation and cooling via H$_2$ down
to temperatures of a few hundreds K. These are not easy conditions to be simultaneously met.  An additional complication in this 
scenario is that a relatively large ($J_{21} > 10^3$ in standard units) Lyman-Werner flux is required which might be found
only locally in some regions of high cosmic star formation. The problem might be somewhat mitigated if collisional dissociation
of H$_2$, as proposed at this conference, can take place efficiently in the cloud interior. All these aspects require further insight
and study.

Feedback by the forming BH, due to its hard radiation, might be a limiting factor for the growth, thus possibly making
the build-up of SMBH in such a brief time truly a challenging problem. Idealized hydrodynamical simulations of feedback-regulated
Bondi accretion have been presented in which it is shown that feedback makes accretion much less efficient than previously
thought. The reason is that the accretion rate is proportional to thermal pressure inside the HII region created by the 
mini-quasar; such pressure prevents material from reaching the bottom of the potential well. Thus, accretion is severely quenched, and
so is the luminosity. After a recombination time of the gas, the process starts again giving rise to a flickering luminosity pattern 
with a typical period equal to the HII region crossing time $\approx R_{HII}/c_s$, where $c_s$ is the sound speed in the ionized
gas. One can conclude that accretion is not continuous even if the BH lives in a relatively rich gas reservoir and should have
a duty cycle set by radiative feedback. Although interesting and novel, these results are drawn from 2D simulations with 
idealized initial conditions. If confirmed by the next generation of studies, they will allow us to clarify in more detail the 
physics of black hole growth.  

\vskip 1.5 truecm
\emph{March 11, 2011 - Dedicated to the memory of those who are not with us anymore.} 






\begin{theacknowledgments}
I would like to thank all the members of the LOC and SOC and all our Japanese friends who have allowed such a
scientifically amazing conference during these harsch and sad times. It is always a pleasure to be back in Japan 
for me, but this time even more so to convey in person all the feelings of friendliness and participation after the 
incredible disaster, which I am sure are shared by the entire world. Many thanks for your proud and organized 
response, I hope it will be an indelible example for all of us.  
\end{theacknowledgments}



\bibliographystyle{aipproc}   

\bibliography{sample}

\IfFileExists{\jobname.bbl}{}
 {\typeout{}
  \typeout{******************************************}
  \typeout{** Please run "bibtex \jobname" to optain}
  \typeout{** the bibliography and then re-run LaTeX}
  \typeout{** twice to fix the references!}
  \typeout{******************************************}
  \typeout{}
 }

\end{document}